\documentclass[aps,prd,preprint,nofootinbib,superscriptaddress]{revtex4}
\usepackage{color}
\usepackage{graphicx}
\usepackage{psfrag}

\begin{document}

\newcommand{\nn}{\nonumber}
\def\dfrac#1#2{\displaystyle\frac{#1}{#2}}
\newcommand{\ovl}[1]{\overline{#1}}
\newcommand{\wt}[1]{\widetilde{#1}}
\newcommand{\eq}[1]{Eq.~(\ref{#1})}
\newcommand{\eqn}[1]{(\ref{#1})}
\newcommand{\p}{\partial}
\newcommand{\pslash}{p\kern-1ex /}
\newcommand{\qslash}{q\kern-1ex /}
\newcommand{\lslash}{l\kern-1ex /}
\newcommand{\sslash}{s\kern-1ex /}
\newcommand{\Dslash}{\mathcal{D}\kern-1.5ex /}
\newcommand{\bpsi}{\overline{\psi}}
\newcommand{\tr}{\mathrm{tr}}
\newcommand{\vev}[1]{\langle #1 \rangle}
\newcommand{\VEV}[1]{\left\langle\mathrm{T} #1\right\rangle}
\newcommand{\rmO}{\mathrm{O}}
\newcommand{\rmd}{\mathrm{d}}
\newcommand{\rme}{\mathrm{e}}
\preprint{UTHEP-675}
\preprint{UTCCS-P-82}
\preprint{RUP-15-9}
\draft

\title{QCD phase transition at real chemical potential with canonical
approach}

\author{Atsushi Nakamura}
\affiliation{
RCNP, Osaka University, Osaka, 567-0047, Japan}
\author{Shotaro Oka}
\affiliation{
Institute of Theoretical Physics, Department of Physics, Rikkyo University,
Toshima-ku, Tokyo 171-8501, Japan}
\author{Yusuke Taniguchi}
\affiliation{
Graduate School of Pure and Applied Sciences, University of Tsukuba,
Tsukuba, Ibaraki 305-8571, Japan}

\begin{abstract}
 We study the finite density phase transition in the lattice QCD at real
 chemical potential.
 We adopt canonical approach and the canonical partition function is
 constructed for $N_f=2$ QCD.
 After derivation of the canonical partition function we calculate
 observables like the pressure, the quark number density, its second
 cumulant and the chiral condensate as a function of the real chemical
 potential.
 We covered a wide range of temperature region starting from the
 confining low to the deconfining high temperature.
 We observe signals for the deconfinement and the chiral restoration
 phase transition at real chemical potential below $T_{c}$ starting from
 the confining phase.
\end{abstract}


\maketitle

\section{Introduction}

The grand canonical ensemble is a difficult subject to treat in lattice
QCD because of the sign problem.
The canonical partition function is related to the grand canonical one
through the fugacity expansion and is known to be free from the complex
action problem.
The history of the canonical approach in finite density lattice QCD may
have started with a development of a reduction formula for the Dirac
determinant \cite{Gibbs:1986hi}, which gives naturally a fugacity
expansion of the Dirac determinant.
A several studies with the staggered fermion are done along this line
\cite{Barbour:1990tb,Hasenfratz:1991ax,Kratochvila:2004wz,deForcrand:2006ec}
and have been taken over by the Wilson fermion
\cite{Nagata:2010xi,Alexandru:2010yb,Li:2011ee,Nagata:2012pc,Nagata:2012tc}.
Since it is understood that the inverse transformation of the fugacity
expansion is the Fourier transformation \cite{Hasenfratz:1991ax} it also
becomes a popular method
\cite{Alexandru:2005ix,Meng:2008hj,Li:2010qf,Danzer:2012vw,Gattringer:2014hra}.

One of the problem of the canonical approach is that the Dirac
determinant needs to be evaluated accurately, which is the heaviest part
of the computation.
One of the solution is to adopt the Taylor expansion
\cite{Ejiri:2008xt}.
In this paper we perform the fugacity expansion by a method of the
hopping parameter expansion in temporal direction: winding number
expansion \cite{Meng:2008hj,Li:2010qf}.
By using the hopping parameter expansion the Dirac determinant can be
evaluated with low cost and we can visit a wide parameter space in
$\beta$.
The canonical partition function is evaluated for $N_f=2$ QCD both in
the deconfinement and the confinement temperature regions.
After derivation of the canonical partition function we study the
chemical potential dependence of observables like the pressure,
the quark number density and the chiral condensate.
We observe signals for the confinement-deconfinement and chiral
restoration phase transition at real chemical potential below the
critical temperature $T_{c}$.
A preliminary result has been reported in Ref.~\cite{Nakamura:2015yga}.

This paper is organized as follows.
In Sec.~2 we briefly review the canonical approach.
Our formulation of the fugacity expansion in terms of the hopping
parameter expansion is discussed in Sec.~3.
After mentioning our numerical setup in Sec.~4 our main results are
given in Sec.~5 for the canonical and grand canonical partition
functions.
Sec.~6 is devoted for the chiral condensate in grand canonical ensemble.
A conclusion is given in Sec.~7.

\section{Canonical partition function}

It is a well known fact that the grand canonical ensemble and the
canonical one are equivalent each other.
This is shown by a simple equation to relate the grand canonical
partition function $Z_G(\mu, T, V)$ and the canonical $Z_C(n, T, V)$
\begin{eqnarray}
Z_G(\mu, T, V)=\sum_{n=-\infty}^\infty Z_C(n,T,V)\xi^n,
\quad
\xi=e^{\mu/T}.
\label{ZG}
\end{eqnarray}
This is a so called fugacity expansion formula.
The inverse of this expansion is given by using the Cauchy's integral
theorem
\begin{eqnarray}
Z_C\left(n,T,V\right)=\oint\frac{d \xi}{2\pi i}\xi^{-n-1}Z_G(\xi,T,V),
\end{eqnarray}
where we assume that the partition function $Z_G(\xi,T,V)$ has
singularities only at $\xi=0, \infty$ corresponding to trivial ones
$\mu/T=\pm\infty$ and adopt an appropriate contour around the origin.
Here we notice that a phase transition point $\xi_c$ is not a
singularity of the partition function but it is rather a zero of
$Z_G(\xi,T,V)$ (Lee-Yang zeros \cite{Yang:1952be,Lee:1952ig}) and does
not affect the Cauchy's integral.

Now it is free to change the contour to a unit circle $\xi=e^{i\theta}$
and the contour integral turns out to be a Fourier transformation
\cite{Hasenfratz:1991ax}
\begin{eqnarray}
Z_C\left(n,T,V\right)=\int_0^{2\pi}\frac{d\theta}{2\pi}
e^{-in\theta}Z_G(e^{i\theta},T,V).
\label{Fourier}
\end{eqnarray}
A standard Monte Carlo simulation is possible for the lattice QCD since
the chemical potential is set to pure imaginary $\mu/T=i\theta$ and
the Boltzmann weight is real positive for two flavors.

However a difficulty of the sign problem is preserved unfortunately
since there remains a frequent cancellation in plus and minus sign in a
numerical Fourier transformation especially for large particle numbers.
In order to get the canonical partition function accurately for large
quark number $n$ we need a very fine sampling of $Z_G(e^{i\theta},T,V)$
in $\theta$.
This requires a heavy computational cost because we need to evaluate the
Dirac operator determinant for the QCD grand partition function.
In this paper we shall solve this problem by a direct expansion of the
Dirac determinant in terms of the fugacity.

\section{Winding number expansion}
\label{sec:WNE}

We consider the lattice $N_f=2$ QCD grand partition function given in
the path integral form
\begin{eqnarray}
Z_G(\xi,T,V)=\int DU \left({\rm Det}D_W(\xi;U)\right)^2e^{-S_G(U)},
\end{eqnarray}
where we adopt the improved Wilson Dirac operator
\begin{eqnarray}
&&
D_W=1-\kappa Q,
\\&&
Q=\sum_{i=1}^3\left(Q_i^++Q_i^-\right)+e^{\mu a}Q_4^++e^{-\mu a}Q_4^-+T,
\\&&
\left(Q_\mu^+\right)_{nm}=\left(1-\gamma_\mu\right)U_\mu(n)
\delta_{m,n+\hat{\mu}},\quad
\left(Q_\mu^-\right)_{nm}=\left(1+\gamma_\mu\right)U_\mu^\dagger(m)
\delta_{m,n-\hat{\mu}},
\\&&
\left(T\right)_{nm}=c_{\rm SW}\frac{1}{2}\sum_{\mu,\nu}\sigma_{\mu\nu}
F_{\mu\nu}(n)\delta_{mn}
\end{eqnarray}

Both the chemical potential $e^{\pm\mu a}$ and the hopping parameter
$\kappa$ appear in front of the temporal hopping term simultaneously.
The fugacity expansion of the determinant shall be given by using the
hopping parameter expansion.

Instead of the determinant we consider the hopping parameter expansion
of
\begin{eqnarray}
{\rm Tr} {\rm Log} D_W={\rm Tr} {\rm Log}\left(I-\kappa Q\right)
=-\sum_{n=1}^\infty\frac{\kappa^n}{n} {\rm Tr}(Q^n).
\label{hpe}
\end{eqnarray}
A non-trivial chemical potential dependence appears when the quark
hoppings make a loop in the temporal direction.
If one of the term has a $n$ times winding loop in positive temporal
direction, the chemical potential dependence is $e^{\mu a nN_t}$ where
$N_t$ is a temporal lattice length and this is nothing but
$e^{n\mu/T}=\xi^n$.
Counting the winding number in temporal direction for each term in
equation (\ref{hpe}), we get the winding number expansion of the TrLog
\begin{eqnarray}
{\rm Tr}{\rm Log}D_W(\mu)=\sum_{n=-\infty}^\infty w_n\xi^n .
\end{eqnarray}
Regrouping the summation, we have a fugacity expansion of the determinant
\cite{Meng:2008hj,Li:2010qf}
\begin{eqnarray}
{\rm Det}D_W(\xi;U)=\exp\left(\sum_{n=-\infty}^\infty w_n\xi^n\right)
=\sum_{n=-\infty}^\infty z_n(U)\xi^n.
\label{regroup}
\end{eqnarray}
In this approach, the fugacity dependence of the determinant or the
partition function is given analytically.
A numerical Fourier transformation (\ref{Fourier}) can be executed
safely at any high precision we want.
In this paper, extraction of the coefficient $z_n(U)$ from the second
term in equation (\ref{regroup}) is done by using the multi-precision
numerical Fourier transformation.

The winding number expansion is done for gauge configurations generated
at $\mu_0=0$ or purely imaginary value.
This procedure corresponds to the standard reweighting method
\begin{eqnarray}
Z_G(\xi,T,V)=\int DU
\left(\frac{{\rm Det}D_W(\mu)}{{\rm Det}D_W(\mu_0)}\right)^2
{\rm Det}D_W(\mu_0)^2e^{-S_G}.
\label{eqn:reweighting}
\end{eqnarray}

\section{Numerical setup}

\begin{figure}
\includegraphics[width=7cm]{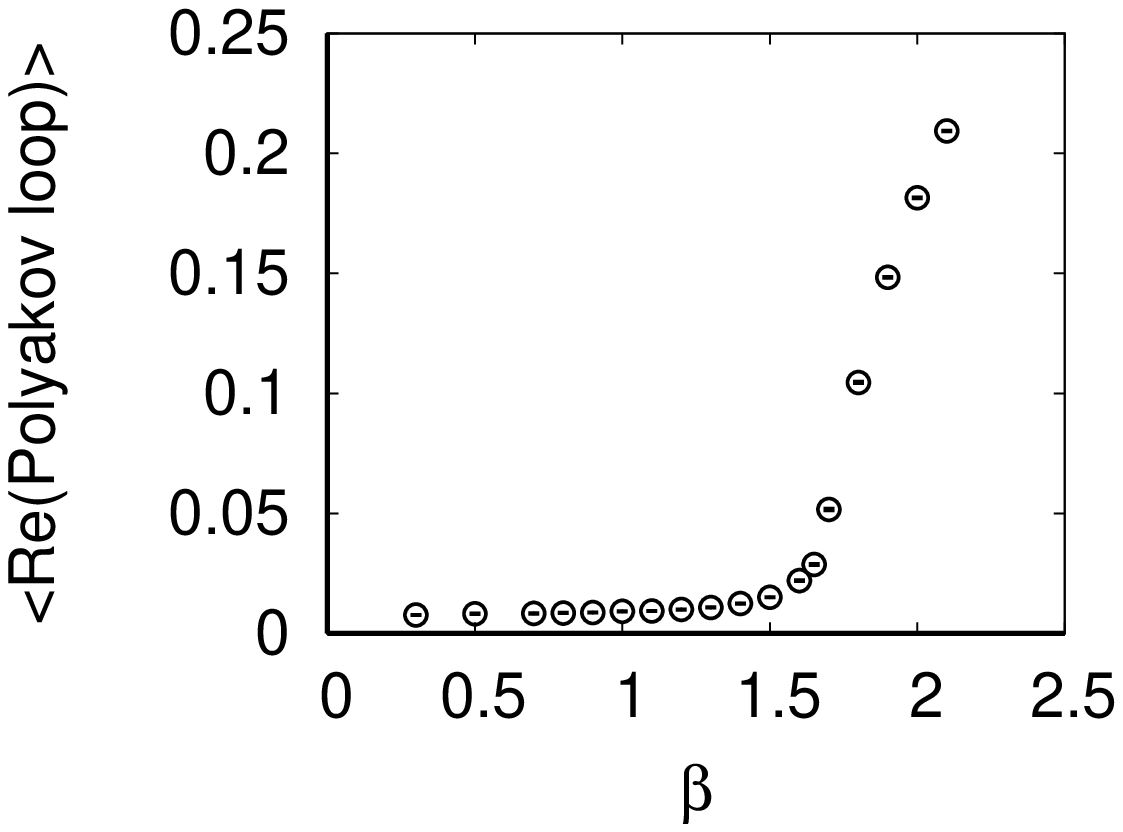}
\includegraphics[width=7cm]{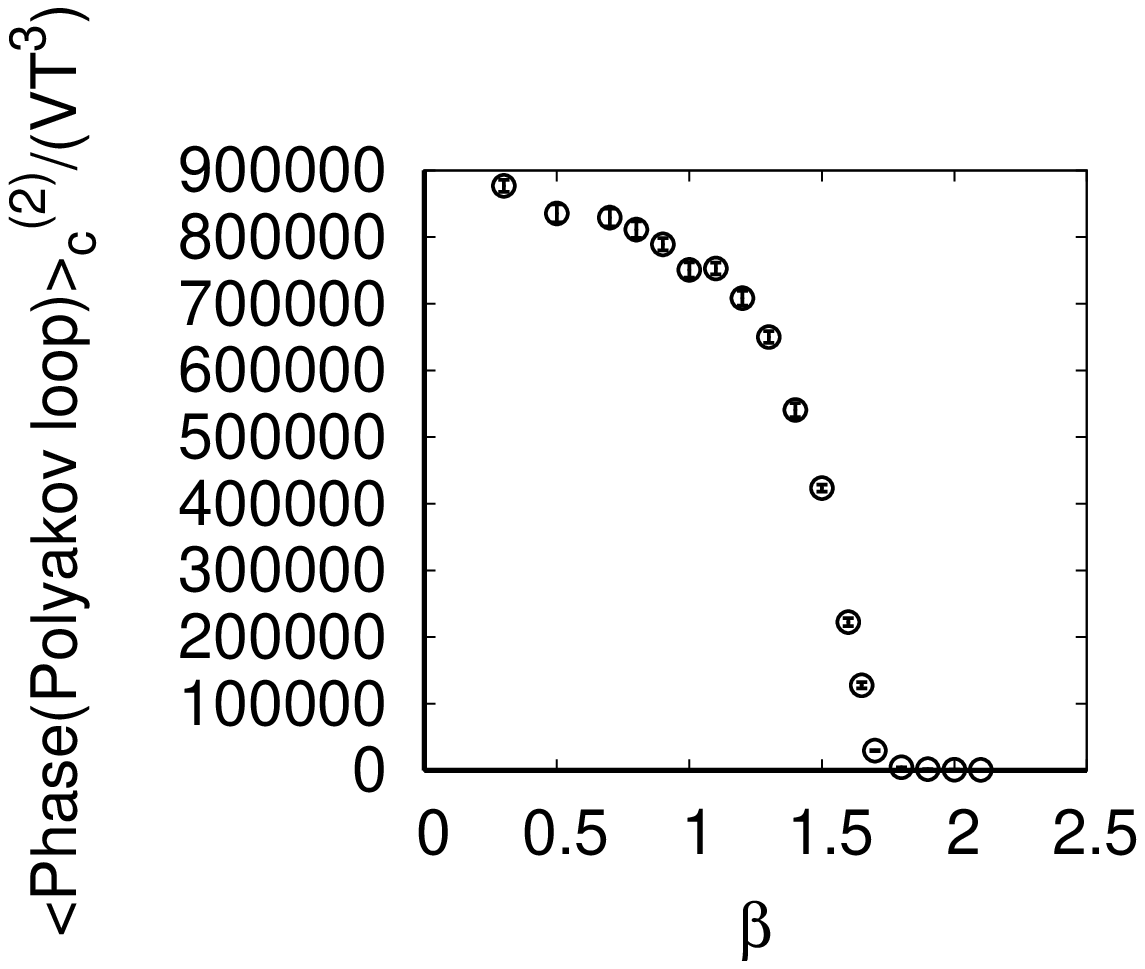}
 \caption{Real part of the Polyakov loop as a function of $\beta$
 (left). Both the confining and deconfining regions are well covered.
 Right panel is a fluctuation of phase of the Polyakov loop.
 $\beta=1.7$ seems to be very near to the critical coupling.}
 \label{fig:polyakov}
\end{figure}

We adopt the Iwasaki gauge action and the improved Wilson fermion action
with the clover term.
The number of flavors is set to $N_f=2$ with degenerate masses.
The APE stout smeared gauge link \cite{Morningstar:2003gk} is used for
those in the fermion action including the clover term.
The number of smearing is four and the parameter is set to $\rho=0.1$.
The clover coefficient is fixed to $c_{\rm SW}=1.1$
\cite{Taniguchi:2013cxa}.
We adopt $8^3\times4$ lattice.
A wide range of $\beta$ is covered from high temperature $\beta=2.1$ to
low temperature $0.9$.
It seems to be that both the confining and deconfining region are well
covered, which is inferred from a behavior of real part of the Polyakov
loop in the left panel of Fig.~\ref{fig:polyakov}.
The temperature corresponding to $\beta=1.7$ seems to be very near to
the critical $T_c$ as is shown the right panel of
Fig.~\ref{fig:polyakov} using a fluctuation of phase of the Polyakov
loop.
The hopping parameter is selected in order that the hopping parameter
expansion works well, for which $m_\pi/m_\rho$ turns out to be $0.7$ -
$0.9$ as is given in table~\ref{table}.
Maximal order of the hopping parameter expansion is set to $480$ so that
max winding number in temporal direction is $120$.
Number of independent configuration is $100$ - $600$.
\begin{table}
\begin{tabular}{|c|c|c|}
\hline
$\beta$ & $\kappa$ & $m_\pi/m_\rho$ \\
\hline
$0.9$ & $0.137$ & $0.8978(55)$\\
$1.1$ & $0.133$ & $0.9038(56)$\\
$1.3$ & $0.138$ & $0.809(12)$\\
$1.5$ & $0.136$ & $0.756(13)$\\
$1.7$ & $0.129$ & $0.770(13)$\\
$1.9$ & $0.125$ & $0.714(15)$\\
$2.1$ & $0.122$ & $0.836(47)$\\
\hline
\end{tabular}
\caption{$\beta$ and $\kappa$ for the numerical simulation.
$m_\pi/m_\rho$ is also given.}
\label{table}
\end{table}

\section{Canonical and grand canonical partition function}

\begin{figure}
 \begin{center}
  \includegraphics[width=7cm]{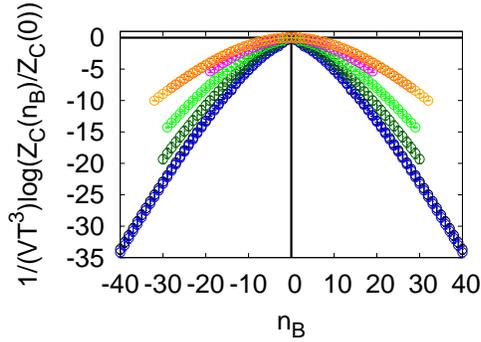}
  \caption{The canonical partition function
  $\log|Z_C(n_B)/Z_C(0)|/(VT^3)$ as a function of the baryon number
  $n_B$.
  From the top $\beta=2.1$ (orange), $1.9$ (red), $1.7$ (magenta), $1.5$
  (green), $1.3$ (dark green), $1.1$ (blue), $0.9$ (black).}
  \label{fig:Zn}
 \end{center}
\end{figure}

The first numerical result we get is the canonical partition function
$Z_C(n,T,V)$.
We plot $\log|Z_C(n_B,T,V)/Z_C(0,T,V)|/(VT^3)$
as a function of the baryon number $n_B$ in Fig.~\ref{fig:Zn}.
The partition function decays very rapidly with $n_B$ and its behavior
changes drastically between $\beta=1.9$ (red) and $1.5$ (green),
which may correspond to a phase transition.

\begin{figure}
 \begin{center}
  \includegraphics[width=7cm]{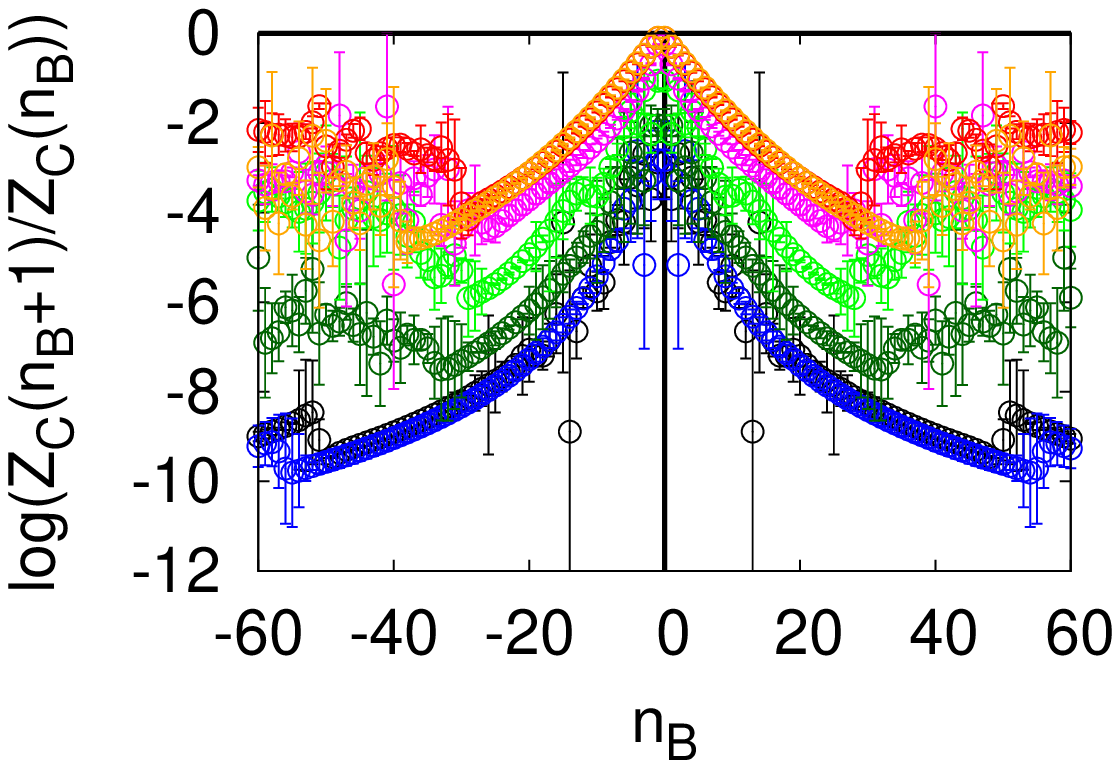}
  \includegraphics[width=7cm]{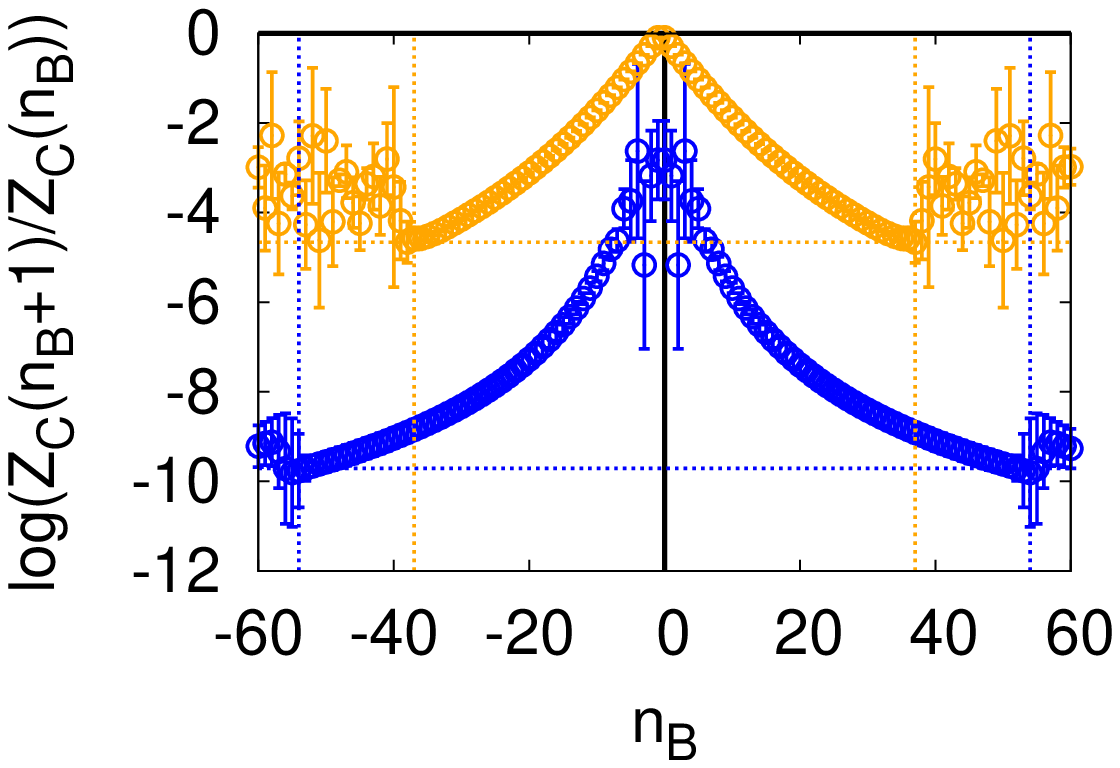}
  \caption{Log plot of $|Z_C(n_B+1)/Z_C(n_B)|$ as a function of the
  baryon number.
  The color and $\beta$ correspondence is the same as in Fig.~2 for left
  panel.
  Right panel is that at $\beta=1.1$ and $2.1$ for example.
}
  \label{fig:dAlembert}
 \end{center}
\end{figure}

The plot range is fixed by using the d'Alembert's convergence condition
\begin{eqnarray}
\lim_{n_B\to\infty}\left|\frac{Z_C(n_B+1)}{Z_C(n_B)}\xi\right|<1,
\end{eqnarray}
which gives the convergence radius for the fugacity $\xi$
\footnote{For negative baryon number we adopt $|Z_C(n_B-1)/Z_C(n_B)|$.}.
The data at $\beta=1.1$ and $2.1$ are plotted in right panel of
Fig.~\ref{fig:dAlembert} for example.
We cut our data at $n_{\rm max}$ where a monotonic decrease stops
indicated by vertical orange and blue lines in the figure.
The horizontal lines show maximal values of the fugacity expected
to be within the convergence radius.
By taking log, the line gives our applicable limit for the baryon
chemical potential $\mu_B/T$.
For example we can discuss physics safely at $-10<\mu_B/T<10$ for
$\beta=1.1$ and $-4<\mu_B/T<4$ for $\beta=2.1$ with our method.
We notice that $n_{\rm max}$ is improved only by one even if we increase
the maximal winding number twice as $120\to240$.
This may be due to a fact that the convergence radius of the Taylor
expansion \eqn{hpe} is unity.

\begin{figure}
 \begin{center}
  \includegraphics[width=7cm]{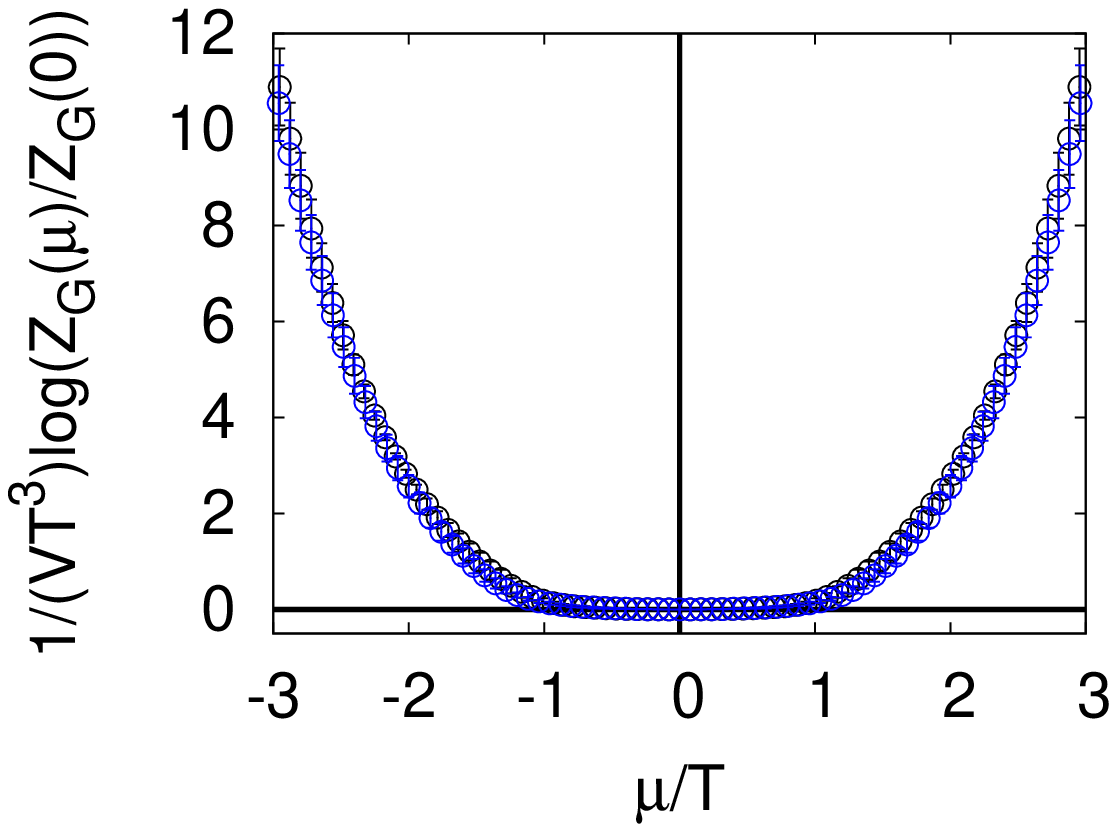}
  \includegraphics[width=7cm]{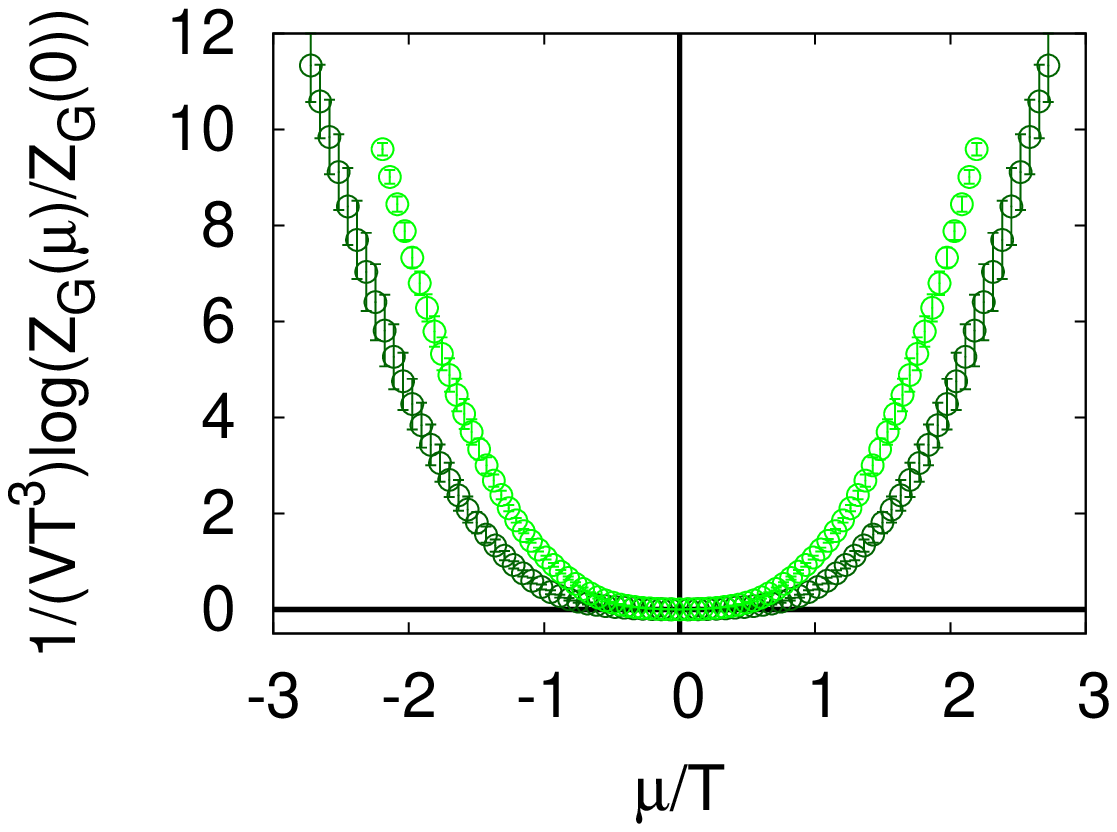}
  \includegraphics[width=7cm]{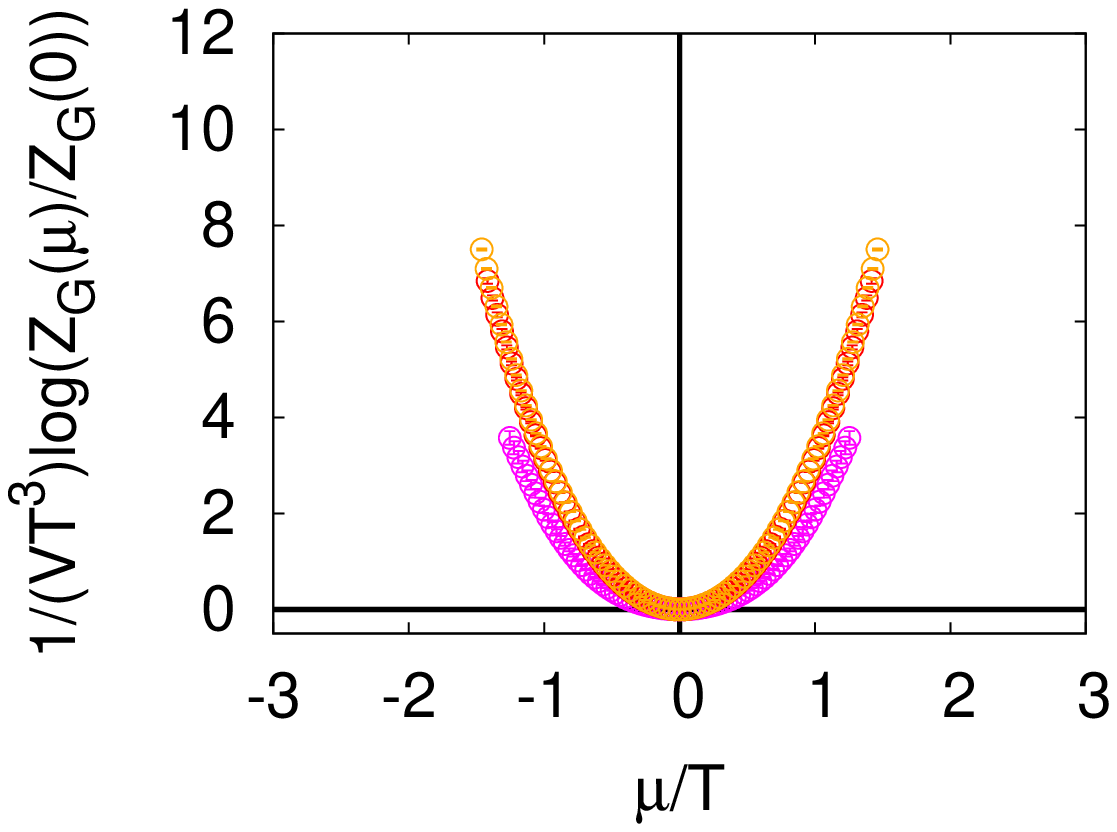}
  \caption{The grand partition function
  $\log\left|Z_G(\mu,T,V)/Z_G(0,T,V)\right|/(VT^3)$ as a function of the
  quark chemical potential $\mu/T$.
  The color and $\beta$ correspondence is the same as in Fig.~2.
  The pressure at $\beta=0.9$ and $1.1$ are plotted in the upper left
  panel, where two data are completely overlapped.
  The upper right panel is that at $\beta=1.3$ and $1.5$.
  Lower panel is that for $\beta=1.7$, $1.9$, $2.1$.}
  \label{fig:ZG}
 \end{center}
\end{figure}

The second physical quantity is the grand partition function.
By taking summation for $-n_{\rm max}\le n_B\le n_{\rm max}$ in
equation (\ref{ZG}), we get the grand partition function for the real
chemical potential.
Here we notice that the canonical partition function $Z_C(n,T,V)$ is a
real positive quantity if a Hermitian transfer matrix exists for the
lattice QCD.
We observed that the absolute value of the partition function is very
stable against the statistical fluctuation and we consider it is
suitable for extracting physical information.
Instead of equation (\ref{ZG}) we adopt 
\begin{eqnarray}
Z_G(\mu, T, V)=\sum_{n=-n_{\rm max}}^{n_{\rm max}} \left|Z_C(n,T,V)\right|\xi^n
\end{eqnarray}
for a definition of the grand partition function
\footnote{Notice this is not a phase quenching of the Dirac
determinant.}, from which we get the
pressure in the grand canonical ensemble
\begin{eqnarray}
\frac{1}{T^4}(P(\mu/T)-P(0))
=\frac{1}{VT^3}\log\left|\frac{Z_G(\mu,T,V)}{Z_G(0,T,V)}\right|.
\end{eqnarray}
We plot the pressure normalized at $\mu=0$ in Fig.~\ref{fig:ZG} as a
function of the quark chemical potential.
We observe that the pressure is almost consistent with zero for small
chemical potential at low temperature below $T_{c}$ as is shown in the
upper panels of Fig.~\ref{fig:ZG}.

\begin{figure}
 \begin{center}
  \includegraphics[width=7cm]{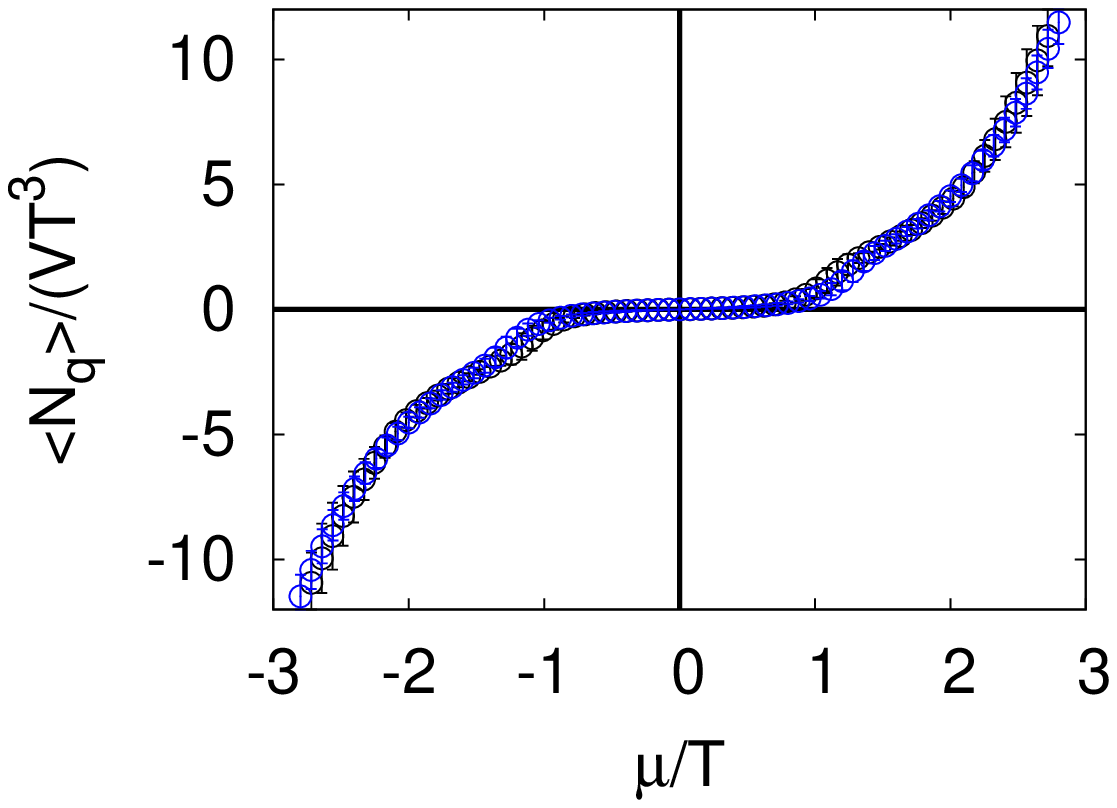}
  \includegraphics[width=7cm]{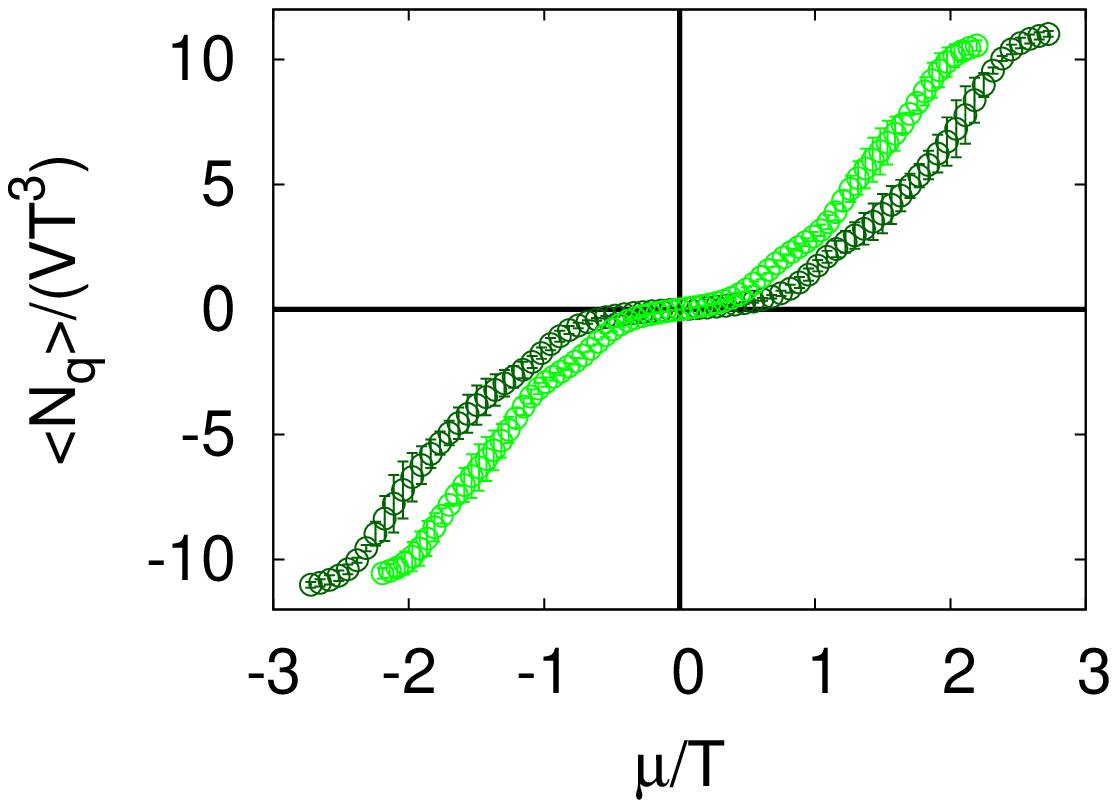}
  \includegraphics[width=7cm]{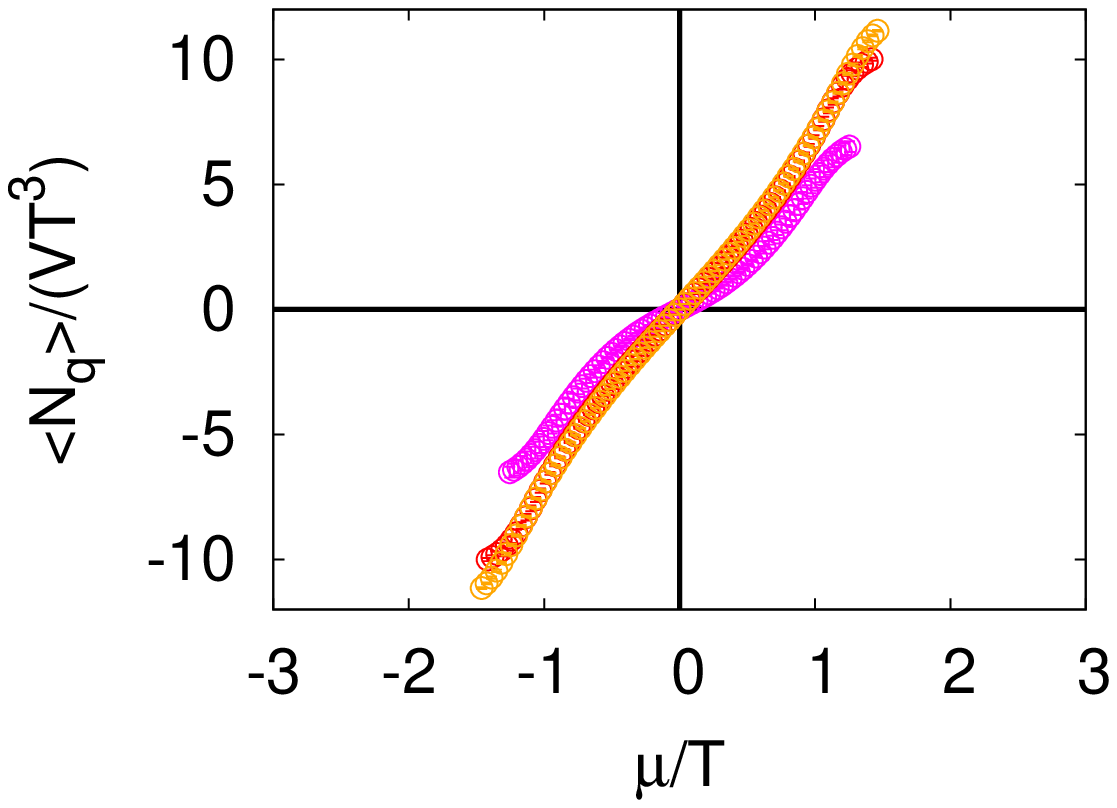}
  \caption{Quark number density $\vev{\hat{N}}/(VT^{3})$ 
  as a function of the quark chemical potential.
  The color and $\beta$ correspondence is the same as in Fig.~2.
  The upper panels plot results for $\beta=0.9$ - $1.5$ below $T_{c}$.
  The lower panel is results for $\beta=1.7$ - $2.1$ above $T_{c}$. }
  \label{fig:nq}
 \end{center}
\end{figure}

The third physical application is the $k$-th momentum of the quark
number operator
\begin{eqnarray}
\vev{\hat{N}^{k}}=\frac{1}{Z_{G}(\mu,T,V)}\sum_{n=-n_{\rm max}}^{n_{\rm max}}
n^{k}Z_C(n,T,V)\xi^n.
\label{Nk}
\end{eqnarray}
In Fig.~\ref{fig:nq}, the quark number density $\vev{\hat{N}}/(VT^{3})$
is plotted as a function of the real quark chemical potential $\mu/T$.
We observe a Silver blaze like phenomenon at low temperature below
$T_{c}$ as is shown in the upper panels.
The quark number density is consistent with zero within a statistical
error until $\mu/T\sim1$ is reached.
Above $\mu/T\sim1$ the quark number density takes off and becomes
non-zero, which indicates deconfinement phase transition.

\begin{figure}
 \begin{center}
  \includegraphics[width=7cm]{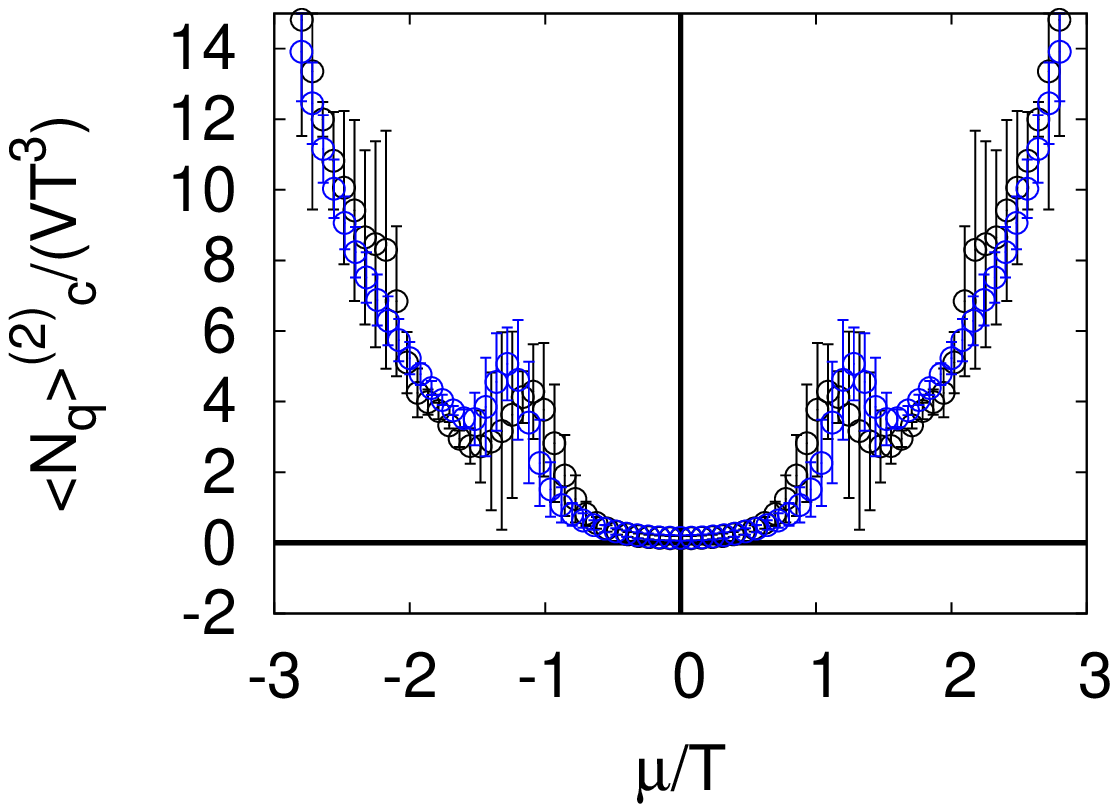}
  \includegraphics[width=7cm]{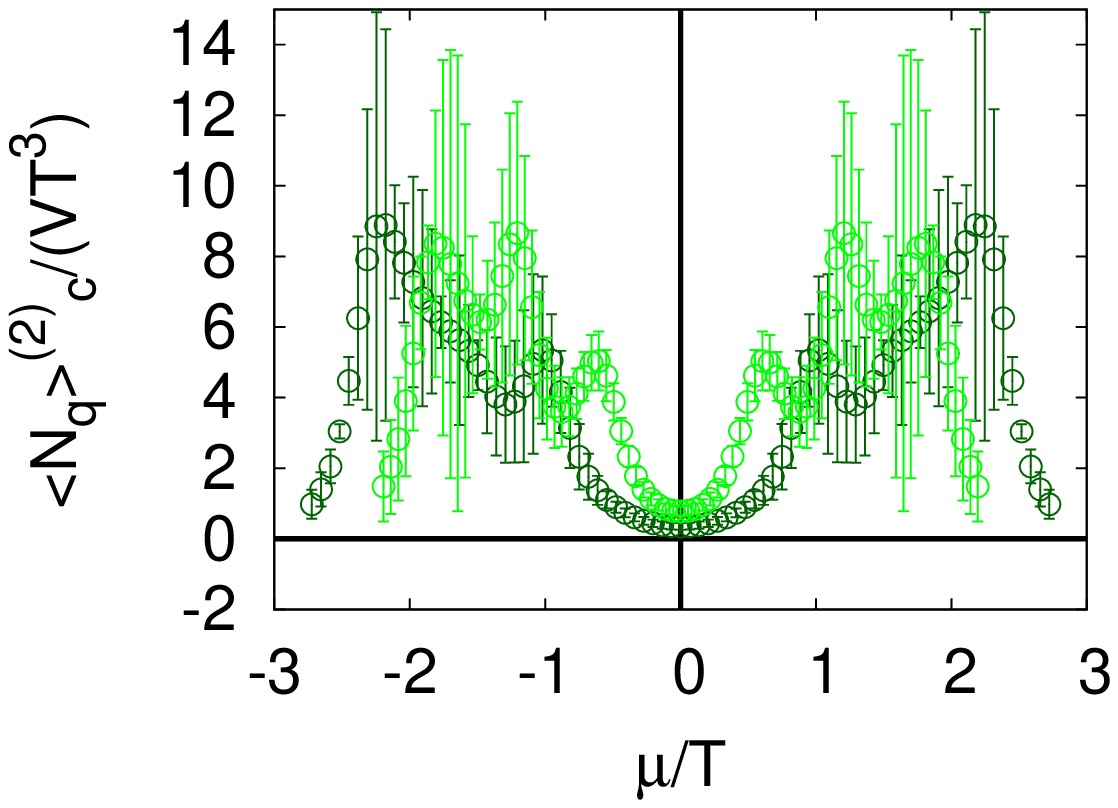}
  \includegraphics[width=7cm]{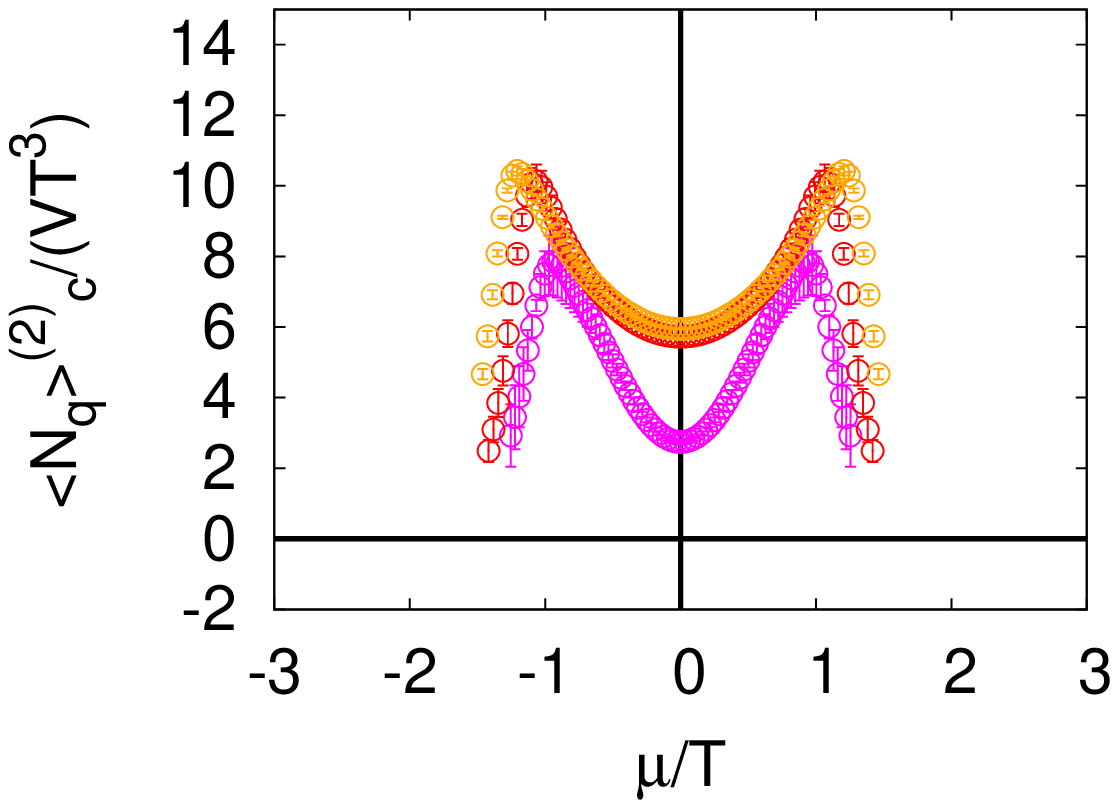}
  \caption{The second cumulant of the quark number
  $\vev{\hat{N}^{2}}_{c}/(VT^{3})$ as a function of the quark chemical
  potential.
  The upper panels plot results for $\beta=0.9$ - $1.5$ below $T_{c}$.
  The lower panel is results for $\beta=1.7$ - $2.1$ above $T_{c}$. }
  \label{fig:nq2}
 \end{center}
\end{figure}

The second cumulant of the quark number density may be a good indicator
of this phase transition.
The results at low temperature side below $T_c$ are plotted in the upper
panels of Fig.~\ref{fig:nq2}.
Those above $T_c$ are given in the lower panel.
We observe peaks around $\mu/T\sim1$ below $T_c$ for every $\beta$ in
the upper panels.
We also observe secondary peaks for $\beta=1.3$ (dark-green) and $1.5$
(green) at larger $\mu/T\geq2$ in the upper right panel.
Only a single peak is observed at high temperature in the lower panel.
These peaks have a distinctive behavior against the quark number cut-off
$n_{\rm max}$.
The first peaks below $T_c$ are stable even if we vary $n_{\rm max}$ by
$20\%$.
On the other hand, the second peaks at $\beta=1.3$, $1.5$ and the first
peaks above $T_c$ change their height and position with $n_{\rm max}$.
It would be conjectured that the first peaks in the low temperature side
are the physical indication of the confinement-deconfinement phase
transition.
The secondary peaks below $T_c$ and the peaks in the high temperature
side would be artifacts due to the cut-off in quark number in the
fugacity expansion.

\begin{figure}
 \begin{center}
  \includegraphics[width=7cm]{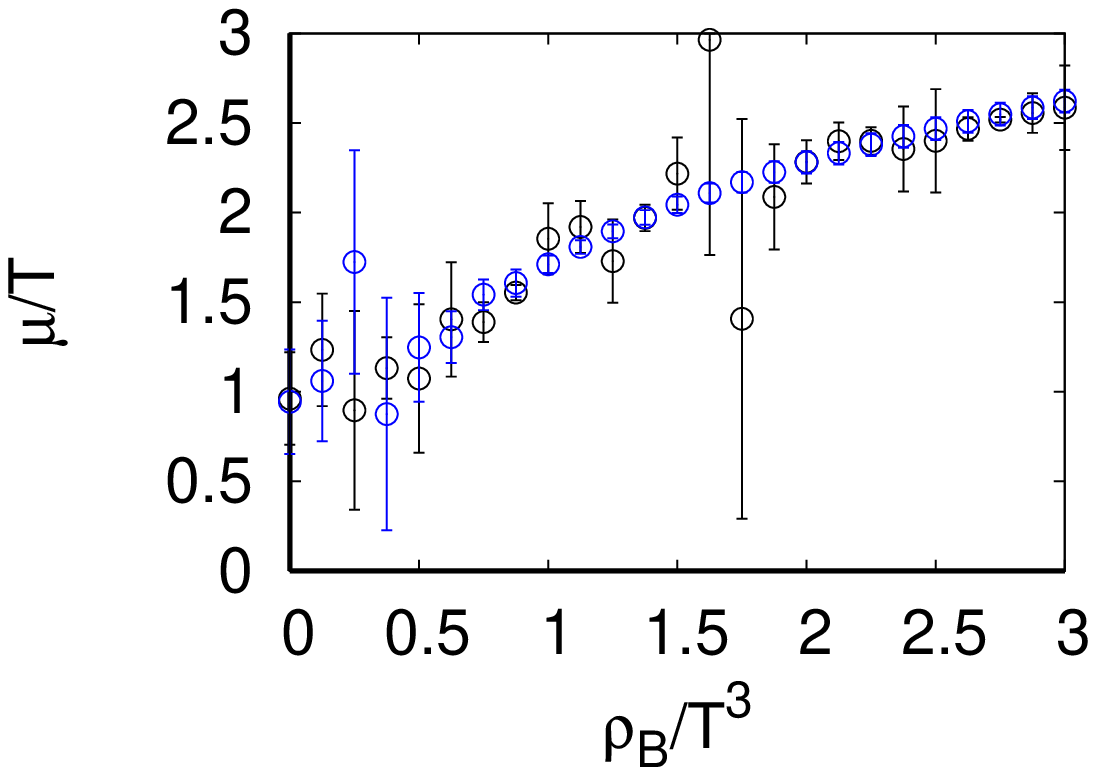}
  \includegraphics[width=7cm]{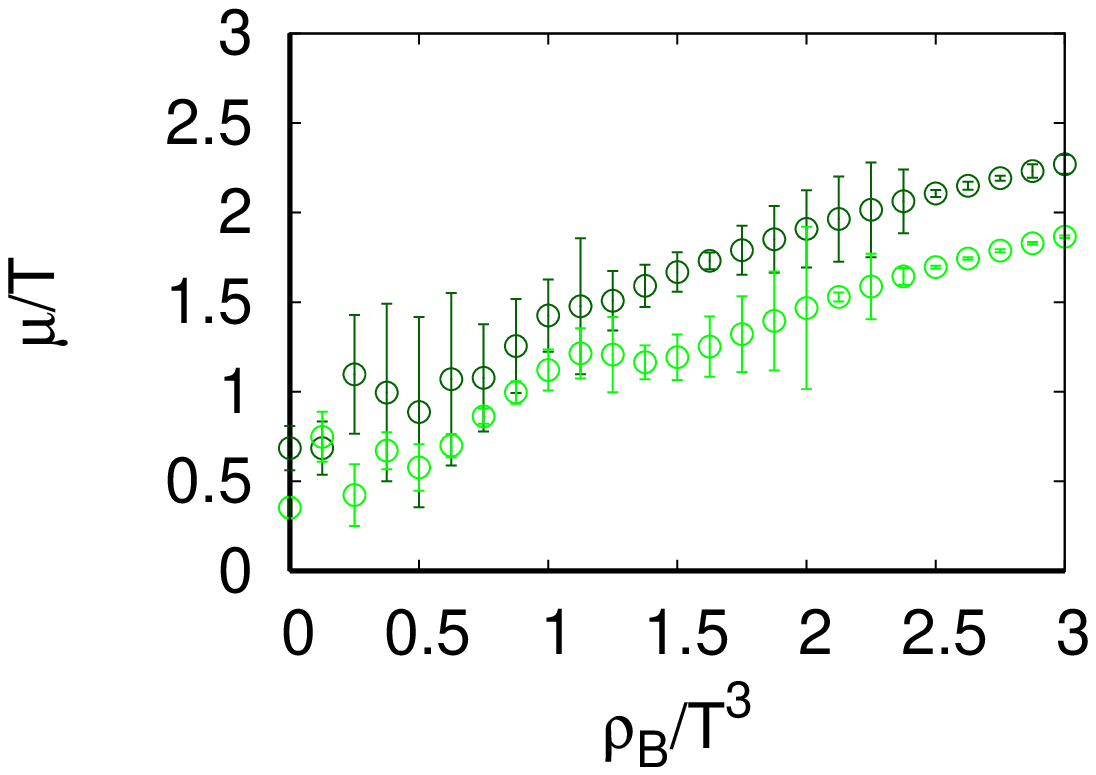}
  \includegraphics[width=7cm]{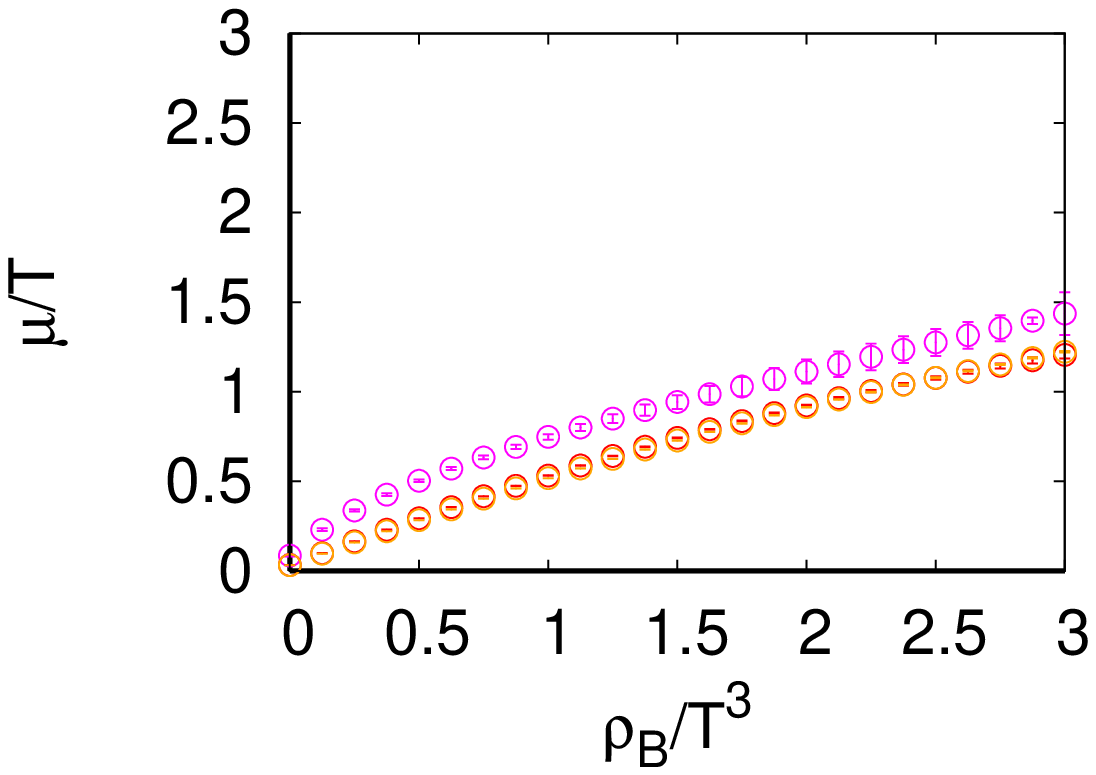}
  \caption{The quark chemical potential $\mu/T$ given in terms of the
  thermodynamical relation as a function of the baryon number density.
  The upper panels plot results for $\beta=0.9$ - $1.5$ below $T_{c}$.
  The lower panel is results for $\beta=1.7$ - $2.1$ above $T_{c}$. }
  \label{fig:mustar}
 \end{center}
\end{figure}

The thermodynamical quantity
\begin{eqnarray}
 \frac{\mu}{T}=-\frac{\p\ln Z_C(n,T,V)}{\p n}
=-\frac{1}{V}\frac{\p\ln Z_C(\rho V,T,V)}{\p \rho}
\label{eqn:mustar}
\end{eqnarray}
as a function of the baryon number density is one of the post popular
quantity in the canonical approach
\cite{deForcrand:2006ec,Li:2011ee,Li:2010qf,Ejiri:2008xt}.
We notice the quantity plotted in Fig.~\ref{fig:dAlembert} can be
interpreted as an expression of the derivative in terms of the
differential
\begin{eqnarray}
 \frac{\mu}{T}=-\frac{1}{3}\left(\ln Z_C(n_B+1,T,V)-\ln Z_C(n_B,T,V)\right).
\end{eqnarray}
The results are plotted in Fig.~\ref{fig:mustar}.
Although the error is rather large there seems to appear a typical
S-shape like behavior at small baryon density $\rho_B/T<0.5$ for the low
temperature side given in the upper panels, which may indicate an
existence of the first order phase transition.

\section{Hadronic observables}

In our procedure, the fugacity expansion is based on the hopping
parameter expansion.
It may be possible to expand any hadronic operators in terms of the
fugacity.
We consider a fugacity expansion of a numerator of some operator vacuum expectation value (VEV)
\begin{eqnarray}
&&
\left\langle O\right\rangle_G(\xi,T,V)=\frac{O_G(\xi,T,V)}{Z_G(\xi,T,V)},
\\&&
O_G(\xi,T,V)=\int DU \left\langle O\right\rangle_{\rm quark}(\xi){\rm Det}D_W(\xi;U)e^{-S_G(U)}
=\sum_{n=-\infty}^\infty O_C(n,T,V)\xi^n,
\label{eqn:OG}
\end{eqnarray}
where $\left\langle O\right\rangle_{\rm quark}$ is a Wick contraction of a hadronic operator $O$
in quark fields.
For example, we consider the chiral condensate.
It is easy to expand its Wick contraction in terms of the hopping parameter
\begin{eqnarray}
\left\langle\bar\psi\psi\right\rangle_{\rm quark}
=-{\rm tr}\left(\frac{1}{D_W}\right)=-{\rm tr}\left(\frac{1}{1-\kappa Q}\right)
=\sum_{m=0}^\infty\kappa^m{\rm tr} Q^m
=\sum_{n=-\infty}^\infty o_n(U)\xi^n.
\end{eqnarray}
Counting the winding number in temporal direction, we get the last equality.
Multiplying the determinant contribution (\ref{regroup}), we apply the
same regrouping procedure in Sec.~\ref{sec:WNE} and get the fugacity
expansion (\ref{eqn:OG}).

\begin{figure}
 \begin{center}
  \includegraphics[width=7cm]{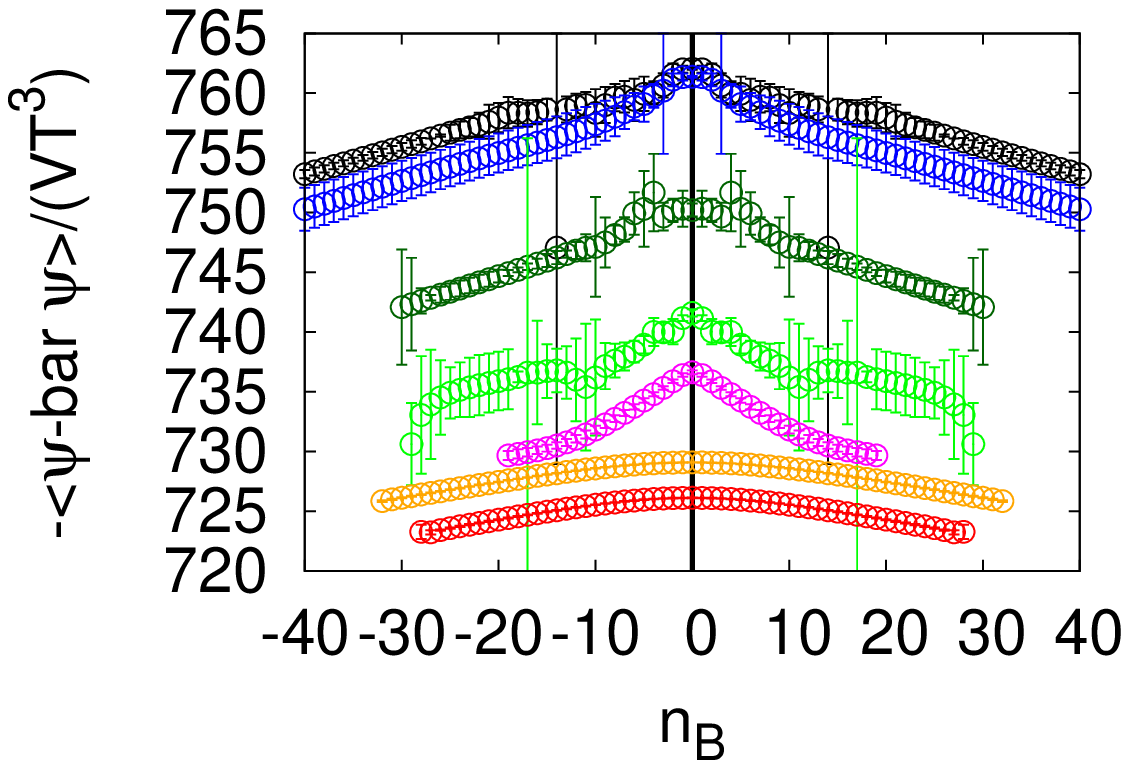}
  \includegraphics[width=7cm]{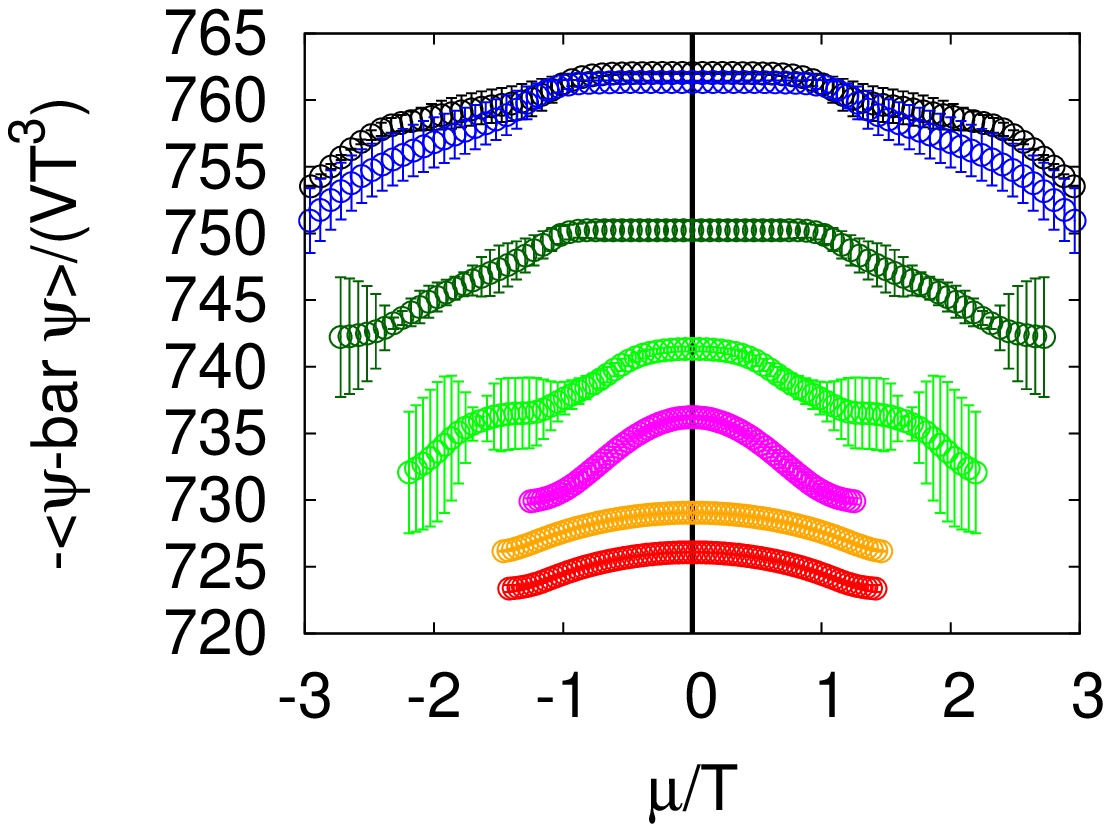}
  \caption{The chiral condensate
  $-\int d^3x\left\langle\bar\psi\psi\right\rangle/(VT^3)$
  in the canonical ensemble (left panel) as a function of the baryon
  number. 
  The right panel is that in the grand canonical ensemble as a function
  of the   quark chemical potential $\mu/T$.
  The color and $\beta$ correspondence is the same as in Fig.~2.
}
  \label{fig:bpsipsi}
 \end{center}
\end{figure}

Once we have two coefficients $Z_C(n,T,V)$ and $O_C(n,T,V)$, the VEVs of
the operator with the canonical and the grand canonical ensemble are
available.
The canonical ensemble VEV is given by taking the ratio of two
coefficient
\begin{eqnarray}
\left\langle O\right\rangle_C(n,T,V)=\frac{O_C(n,T,V)}{Z_C(n,T,V)}.
\end{eqnarray}
The result for the chiral condensate
$-\int d^3x\left\langle\bar\psi\psi\right\rangle_C/(VT^3)$
is given in the left panel of Fig.~\ref{fig:bpsipsi} as a function of
the baryon number.

A VEV in the grand canonical ensemble is given by taking fugacity
summation with real chemical potential
\begin{eqnarray}
\left\langle O\right\rangle_G(\xi,T,V)
=\frac{\sum_{n=-n_{\rm max}}^{n_{\rm max}}O_C(n,T,V)\xi^n}
{\sum_{n=-n_{\rm max}}^{n_{\rm max}}Z_C(n,T,V)\xi^n}.
\end{eqnarray}
The chiral condensate
 $-\int d^3x\left\langle\bar\psi\psi\right\rangle_G/(VT^3)$ is
given in the right panel of Fig.~\ref{fig:bpsipsi} as a function of the
real chemical potential $\mu/T$.
The condensate in the figure is a bare quantity without renormalization.
Since we adopted the Wilson fermion, we have an additive correction for
$\left\langle\bar\psi\psi\right\rangle$, which is not subtracted in this
paper.
From the figure, the chiral restoration phase transition at finite
chemical potential seems to be seen.
A relatively large value at around $\mu/T=0$ starts to decrease rapidly
at chemical potential $\mu/T\sim1$ for low temperature.
The would-be transition parameter $\mu_c/T$ becomes larger for lower
temperature and smaller for high temperature although we need to notice
the quark mass is not the same for each $\beta$.
The position of $\mu_c/T$ seems to be consistent with that of the first
peak for the second cumulant in the upper panels of Fig.~\ref{fig:nq2}.

\begin{figure}
 \begin{center}
  \includegraphics[width=7cm]{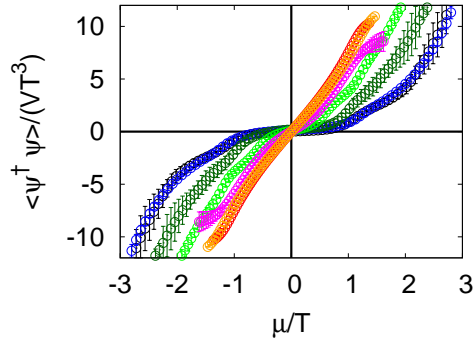}
  \caption{The quark number density
  $\int d^3x\left\langle\psi^\dagger\psi\right\rangle/(VT^3)$
  in the grand canonical ensemble as a function of the
  quark chemical potential $\mu/T$.
  The color and $\beta$ correspondence is the same as in Fig.~2.
}
  \label{fig:psipsi}
 \end{center}
\end{figure}

The quark number density is also given in terms of the hadronic
observable
$\int d^3\left\langle\psi^\dagger\psi\right\rangle/(VT^3)$, which gives
consistent result with that in Sec.~5 up to renormalization factor as is
plotted in Fig~\ref{fig:psipsi}.

\section{Conclusion}

In this paper, we performed the fugacity expansion of the grand partition
function by using the hopping parameter expansion.
This procedure seems to be valid for baryon numbers around $n_B\sim30$
for $8^3\times4$ lattice and $m_\pi/m_\rho>0.7$.
The method is also applied to the numerator of VEVs of hadronic
operators.
Taking summation over quark numbers we get a VEV at the real chemical
potential.
As an example we evaluate the pressure, the quark number density and its
second cumulant and the chiral condensate.
These observables show a phase transition like behavior at high chemical
potential for low temperature region, which may correspond to the
confinement-deconfinement or the chiral restoration phase transition.

One of the biggest anxiety in this paper may be an overlap problem due to
the reweighting \eqn{eqn:reweighting}.
We performed a test of this problem by adopting $\mu_0/T=0.5$ for
configurations generation.
We find all the results are consistent within the statistical error as
we shall report in the forthcoming paper.

\section*{Acknowledgement}

We would like to thank T.~Eguchi, R.~Fukuda, M.~Harada, T.~Matsumoto,
S.~Sakai, Y.~Shimizu, A.~Suzuki, H.~Suzuki, M.~Yahiro for valuable
discussions.
This work is done for Zn Collaboration.
This work is supported in part by Grants-in-Aid of the Ministry of
Education (Nos. 26610072, 24340054, 22540265, 15H03663, 26610072).
This work is in part based on Bridge++ code
(http://suchix.kek.jp/bridge/Lattice-code/).

\end{document}